\documentclass[pra,showpacs,preprintnumbers, amsmath,amssymb,floatfix]{revtex4}
\usepackage{hyperref}
\usepackage[dvips]{epsfig}
\newcommand{\beq}{\begin{equation}}
\newcommand{\eeq}{\end{equation}} 
\newcommand{\beqa}{\begin{eqnarray}}
\newcommand{\eeqa}{\end{eqnarray}}
\newcommand{\ba}{\begin{array}}
\newcommand{\ea}{\end{array}}

\begin{document}

\title{Hydrodynamics of Bose and Fermi superfluids at zero temperature: \\
the superfluid nonlinear Schr\"odinger equation} 
\author{Luca Salasnich} 
\affiliation{CNR-INFM and CNISM, Unit of Padua, \\ 
Department of Physics ``Galileo Galilei'', 
University of Padua, \\
Via Marzolo 8, 35122 Padua, Italy} 

\begin{abstract} 
We discuss the zero-temperature hydrodynamics equations 
of bosonic and fermionic superfluids and their connection with 
generalized Gross-Pitaevskii and Ginzburg-Landau equations 
through a single superfluid nonlinear Schr\"odinger equation. 
\end{abstract}


\maketitle

Recent and less-recent experiments with ultracold and dilute gases 
made of alkali-metal atoms have clearly shown the existence 
of superfluid properties in these systems \cite{stringa1,stringa2}. 
Both bosonic and fermionic superfluids can be accurately 
described by the hydrodynamics equations 
of superfluids \cite{stringa1,stringa2,landau1}. 
In this paper we analyze the hydrodynamics equations 
of superfluids and show how to construct a reliable nonlinear 
Schr\"odinger equation from these hydrodynamics equations. 
For bosons this equation gives the generalized Gross-Pitaevskii 
equation recently discussed by Volovik \cite{volovik}, 
while for fermions one gets a zero-temperature 
Ginzburg-Landau equation \cite{sala-gle}. 
The limits of validity of these mean-field equations are discussed. 

At zero temperature the hydrodynamics equations of superfluids made of atoms 
of mass $m$ are given by 
\beqa 
{\partial \over \partial t} n &+& 
\nabla \cdot \left( n {\bf v} \right) = 0 \; , 
\label{hy-1}
\\
m {\partial \over \partial t} {\bf v} &+& 
\nabla \left[ {1\over 2} m v^2 + U({\bf r}) + \mu(n) \right] = 0 \; ,  
\label{hy-2}
\eeqa 
where $n({\bf r},t)$ is the local density and ${\bf v}({\bf r},t)$ 
is the local superfluid velocity \cite{stringa1,stringa2,landau1}. 
Here $U({\bf r})$ is the external potential and $\mu(n)$ 
is the bulk chemical potential of the system. 
The bulk chemical potential $\mu(n)$ is the 
zero-temperature equation of state 
of the uniform system. The density $n({\bf r},t)$ is such that 
\beq 
N = \int n({\bf r},t) \ d^3{\bf r} 
\eeq
is the total number of atoms in the fluid. 
Eq. (\ref{hy-1}) and (\ref{hy-2}) are nothing else than  
the Euler equations of an inviscid (i.e. not-viscous) and 
irrotational fluid. In fact, at zero temperature, due to the 
absence of the normal component, the superfluid density 
coincides with the total density and the superfluid current 
with the total current \cite{landau1}. 

The condition of irrotationality 
\beq 
\nabla \wedge {\bf v} = 0 
\label{irr}
\eeq
means that the velocity ${\bf v}$ can be written as the 
gradient of a scalar field. Eqs. (\ref{hy-1}) and (\ref{hy-2}) 
differ from the corresponding equations holding in the 
collisional regime of a non superfluid system because of 
the irrotationality constraint (\ref{irr}). 
In addition, experiments with both 
bosonic and fermionic superfluids show the existence 
of quantized vortices, such that the circulation ${\cal C}$ 
of the superfluid velocity 
\beq 
{\cal C} = \oint {\bf v} \cdot d{\bf r} \; , 
\eeq 
is quantized, i.e. 
\beq 
{\cal C} = {2\pi \hbar \ k \over \zeta m} \; , 
\label{quantized}
\eeq 
where $k$ is an integer quantum number and the statistical 
coefficient $\zeta$ is $1$ for superfluid bosons and 
$2$ for superfluid fermions \cite{stringa1,stringa2,landau2,leggett}. 

Eq. (\ref{quantized}) does not have a classical analog, 
and this fact suggests that the superfluid 
velocity is the gradient of the phase $\theta({\bf r},t)$ of 
a single-valued quantum-mechanical wave function 
$\Xi({\bf r},t)$. This function 
\beq 
\Xi({\bf r},t) =|\Xi({\bf r},t)| \ e^{i\theta({\bf r},t)} 
\eeq 
is the so-called macroscopic wave function of the 
Bose-Einstein condensate of the superfluid \cite{landau2,leggett}. 
The connection between superfluid hydrodynamics and 
quantum field theory is made by the formula 
\beq  
{\bf v} = {\hbar \over \zeta m} \nabla \theta \; , 
\label{v-general} 
\eeq 
where again $\zeta=1$ for bosons and $\zeta=2$ for fermions. 
For bosons the condensate wave function is given by 
\beq 
\Xi({\bf r},t) = 
\langle{\hat \psi}({\bf r},t)\rangle \; , 
\eeq
that is the thermal (or ground-state) average 
of the bosonic field operator 
${\hat \psi}({\bf r},t)$ \cite{landau2}. 
For fermions the condensate wave function is instead 
\beq 
\Xi({\bf r},t) = 
\langle {\hat \psi}_{\uparrow}({\bf r},t) 
{\hat \psi}_{\downarrow}({\bf r},t) \rangle \; , 
\eeq 
that is the average of pair operators, 
with ${\hat \psi}_{\sigma}({\bf r},t)$ the fermionic 
field operator with spin component $\sigma=\uparrow,\downarrow$ 
\cite{landau2}. Notice that the condensate wave function 
$\Xi({\bf r},t)$ is normalized to the total number 
of condensed bosons \cite{landau2} or of condensed Cooper pairs 
\cite{leggett,sala-odlro}, and in general the condensed fraction 
can be much smaller than one \cite{landau2,sala-odlro}. 

The effect of statistics enters in the explicit formula 
\cite{stringa1,stringa2,sala-ideal}
of the bulk chemical potential $\mu(n)$ which appears 
in Eqs. (\ref{hy-1}) and (\ref{hy-2}). For instance, 
in the case of a dilute and ultracold 
wekly-interacting Bose gas one has 
\beq 
\mu(n) = {4 \pi \hbar^2 \over m} n 
\left( 1 + {32\over 3\sqrt{\pi}} a_B n^{1/3} \right) 
\; ,
\label{mu-bose}
\eeq
where $a_B$ is the inter-atomic Bose-Bose 
s-wave scattering length and $a_B n^{1/3}$ 
is the gas parameter of the bosonic Lee-Yang expansion \cite{lee}. 
Very recently we have proposed a Pad\`e approximant 
to describe $\mu(n)$ from the the weak-coupling regime, 
where $a_B n^{1/3}\ll 1$ and Eq. (\ref{mu-bose}) holds, to 
the unitarity limit, where $a\to +\infty$ \cite{sala-ggpe}. 
In the case of a dilute and ultracold two-component 
weakly-interacting Fermi gas one can take instead 
\beq 
\mu(n) = {\hbar^2\over 2m} \left(3\pi^2 n\right)^{2/3} 
\left( 1 + {4\over 3\pi} \left(3\pi^2\right)^{1/3} a_F n^{1/3} 
\right) 
\; ,      
\label{mu-fermi} 
\eeq 
where $a_F$ is the inter-atomic 
scattering length of atoms with different spin 
and $a_F n^{1/3}$ is the gas parameter 
of the fermionic Huang-Yang expansion \cite{huang}. 
Few years ago we proposed a reliable fitting 
formula of $\mu(n)$ based on Monte Carlo data 
in the full BCS-BEC crossover from weakly-interacting Cooper pairs, 
where Eq. (\ref{mu-fermi}) holds, 
to the Bose-Einstein condensate of molecules \cite{manini05}. 

The hydrodynamics equations (\ref{hy-1}) 
and (\ref{hy-2}) are valid to describe the long-wavelength and 
low-energy macroscopic properties of both bosons and fermions. 
In particular, one can introduce a healing (or coherence) 
length $\xi$ such that the phenomena under investigation 
must be characterized by a wave length $\lambda$ much larger than 
the healing length, i.e. 
\beq 
\lambda \gg \xi \; . 
\eeq 
As suggested by Combescot, Kagan and Stringari \cite{combescot}, 
the healing length can be defined as 
\beq 
\xi = {\hbar\over m v_{cr}} \; , 
\label{healing}
\eeq
where $v_{cr}$ is the Landau critical velocity above 
which the system gives rise to energy dissipation. 
For bosons the critical velocity coincides 
with the first sound velocity, i.e. 
\beq 
v_{cr} = \sqrt{{n\over m}{\partial \mu \over \partial n}} \; . 
\label{v-cr-b}
\eeq
For fermions the critical velocity is instead 
related to the breaking of Cooper pairs through the formula 
\beq 
v_{cr} = \sqrt{ \sqrt{\mu^2 + |\Delta|^2}-\mu \over m} \; , 
\label{v-cr-f}
\eeq
where $|\Delta|$ is the energy gap of Cooper pairs \cite{stringa2,combescot}. 
We notice that in the deep BCS regime of weakly interacting 
attractive Fermi atoms 
(corresponding to $|\Delta| \ll \mu$) Eq. (\ref{v-cr-f}) 
approaches the exponentially small value $v_{cr}=|\Delta|/\sqrt{2m \mu}$. 
In addition, we remind that some years ago 
Kemoklidze and Pitaevskii derived the zero-temperature 
Eqs. (\ref{hy-1}) and (\ref{hy-2}) for the BCS Fermi gas 
starting from the Gorkov equations 
of quantum-field theory under the assumption of neglecting 
the spatial derivatives of the energy gap $|\Delta|$ \cite{kemo}. 
In the case of a Fermi gas that performs the whole BCS-BEC crossover 
the critical velocity $v_{cr}$ can be estimated as the minimum 
value between the two values given by Eq. (\ref{v-cr-b}) and 
Eq. (\ref{v-cr-f}) \cite{combescot}. 

Inspired by the Ginzburg-Landau theory of superconductors 
\cite{ginzburg} and by the density functional approach 
to the superfluid $^4$He \cite{density}, 
we now try to express Eqs. (\ref{hy-1}) and (\ref{hy-2}) 
in terms of a Schr\"odinger equation such that 
its wave function $\Psi({\bf r},t)$ is given by 
\beq
\Psi({\bf r},t) = \sqrt{n({\bf r},t)\over \zeta} 
\ e^{ i \theta({\bf r},t) } \; ,  
\label{psi} 
\eeq 
where the coefficient $\zeta$ is $1$ for bosons and 
$2$ for fermions. We call $\zeta$ the statistical coefficient. 
In this way the function $\Psi({\bf r},t)$ 
describes superfluid bosons ($\zeta=1$) or boson-like 
Cooper pairs ($\zeta =2$) with the normalization
\beq 
\int |\Psi({\bf r},t)|^2 d^3{\bf r} = {N\over \zeta} \; ,  
\label{norma} 
\eeq 
that is quite different from the normalization of the 
condensate wave function $\Xi({\bf r},t)$ \cite{landau2,sala-odlro}. 
Nevertheless, the phase $\theta({\bf r},t)$ of the the complex field 
$\Psi({\bf r},t)$ is the same of 
$\Xi({\bf r},t)$ and also of the gap function $\Delta({\bf r},t)$. 
Obviously this phase must satisfy Eq. (\ref{v-general}). 

We consider the following nonlinear Schr\"odinger equation 
\beq
i \hbar {\partial \over \partial t} \Psi 
= \left[ -\alpha {\hbar^2 \over 2m} \nabla^2 + 
\beta \ U + \gamma \ \mu(n) \right] \Psi \; , 
\label{nlse}
\eeq 
where $\alpha$, $\beta$ and $\gamma$ 
are parameters which must be determined.   
By inserting Eq. (\ref{psi}) into Eq. (\ref{nlse}), 
after some calculations and taking into account 
Eq. (\ref{v-general}), we find two hydrodynamics equations
\beqa 
{\partial \over \partial t} n 
&+& \alpha \ \zeta \ \nabla \cdot \left( n {\bf v} \right) = 0 \; , 
\\ 
m {\partial \over \partial t} {\bf v} &+&
\nabla \left[ -{\alpha \over \zeta} 
{\hbar^2\over 2m} 
{\nabla^2 \sqrt{n}\over \sqrt{n}} + 
{\alpha \ \zeta \over 2} m v^2 + {\beta\over \zeta} \ U + 
{\gamma\over \zeta} \ \mu(n) \right] = 0 \; ,  
\eeqa 
which include the quantum pressure term 
\beq 
T_{QP}=-{\alpha \over \zeta}
{\hbar^2\over 2m}
{\nabla^2 \sqrt{n}\over \sqrt{n}} \; , 
\eeq 
which depends explicitly on the reduced Planck constant $\hbar$. 
This term is necessary in a realistic superfluid model to avoid unphysical 
phenomena like the formation of wave front singularities 
in the dynamics of shock waves \cite{sala-shock}.  
Neglecting the quantum pressure term (which is small for a large 
number of particles apart very close the surface), 
one gets the classical hydrodynamics equations, i.e. 
Eqs. (\ref{hy-1}) and (\ref{hy-2}), only by setting 
\beq 
\alpha = {1\over \zeta} \; , \quad\quad  
\beta = \gamma = \zeta \; . 
\eeq
In conclusion Eq. (\ref{nlse}) becomes 
\beq 
i \hbar {\partial \over \partial t} \Psi({\bf r},t) 
= \left[ -{\hbar^2 \over 2\zeta m} \nabla^2 + 
\zeta U({\bf r}) + \zeta \mu(n({\bf r},t)) \right] \Psi({\bf r},t) \; ,
\label{super}
\eeq 
that we call {\it superfluid nonlinear Schr\"odinger} (SNLS) equation, 
and the quantum pressure term reads 
\beq 
T_{QP}=-{\hbar^2\over 2 \zeta^2 m }
{\nabla^2 \sqrt{n}\over \sqrt{n}} \; .
\eeq  
For superfluid bosons $\zeta =1$ and from Eq. (\ref{super}) 
we get a generalized Gross-Pitaevskii equation 
\cite{volovik,berloff,sala-ggpe} which becomes 
the familiar Gross-Pitaevskii equation 
if $\mu(n)=(4\pi\hbar^2 a_B/m)n$ \cite{gross}.  
For superfluid fermions one has instead $\zeta = 2$ 
and from Eq. (\ref{super}) we get a zero-temperature generalized 
Ginzburg-Landau equation \cite{sala-gle,nota}. 
We stress that the appearance of the statistical coefficient 
$\zeta$ in Eq. (\ref{super}) is a direct consequence 
of the relationship (\ref{v-general}) between the phase 
and the superfluid velocity, while the normalization (\ref{norma}) 
does not affect Eq. (\ref{super}). 

It is important to stress that Eq. (\ref{nlse}) is not the 
more general Galilei-invariant Sch\"orodinger equation. In fact, 
we have recently shown \cite{flavio} that 
a nonlinear Schr\"odinger equation of the Guerra-Pusterla type \cite{guerra},  
\beq
i \hbar {\partial \over \partial t} \Psi 
= \left[ -\alpha {\hbar^2 \over 2m} \nabla^2 + 
\beta \ U + \gamma \ \mu(n) + \eta \ {\hbar^2\over 2m} 
{\nabla^2 |\Psi|\over |\Psi|}
\right] \Psi \; , 
\label{nlse-new}
\eeq 
is needed to accurately describe the surface effects of a ultracold 
superfluid Fermi gas with infinite scattering length 
in a harmonic trap \cite{flavio}. 

Eq. (\ref{super}) is the Euler-Lagrange 
equation of the following Lagrangian density 
\beq 
{\cal L} = {i\hbar\over 2} \left(
\Psi^* {\partial \over \partial t} \Psi 
- \Psi {\partial \over \partial t} \Psi^* \right) 
+ {\hbar^2\over 2 \zeta m} \Psi^* \nabla^2 \Psi 
- \zeta U({\bf r}) |\psi|^2 - \zeta  {\cal E}(n) |\Psi|^2 \; , 
\eeq
where 
\beq 
{\cal E}(n) = {1\over n} \int_0^n \mu(n') \ dn' \; 
\eeq
is the bulk energy per particle of the bosonic superfluid, namely 
\beq 
\mu(n)={\partial \left( n{\cal E}(n) \right)\over \partial n} \; . 
\eeq

Our SNLS equation (\ref{super}) can be used to study 
stationary configurations and elementary 
excitations of the superfluids. To obtain 
the stationary equation we set  
\beq 
\Psi({\bf r},t) = \Psi_{eq}({\bf r}) \ e^{-i\zeta \bar{\mu} t/\hbar} \; , 
\eeq 
where $\Psi_{eq}({\bf r})$ is the equilibrium wave function 
which satisfies the stationary equation 
\beq 
\left[ -{\hbar^2 \over 2 \zeta m} \nabla^2 + 
\zeta U({\bf r}) + \zeta  \mu(n_{eq}({\bf r})) \right] 
\Psi_{eq}({\bf r}) = \zeta  \bar{\mu}\ \Psi_{eq}({\bf r}) \; ,
\eeq 
with $n_{eq}({\bf r})=\zeta |\Psi_{eq}({\bf r})|^2$ the stationary 
density profile of superfluid bosons ($\zeta=1$) 
or fermions ($\zeta=2$), and $\bar{\mu}$ the chemical potential 
of the inhomogeneous system. 
Neglecting the gradient term (Thomas-Fermi approximation) 
from the stationary SNLS equation 
we obtain the same algebric equation for both bosons and fermions, 
namely 
\beq 
U({\bf r}) + \mu(n_{eq}({\bf r})) = \bar{\mu} \; ,    
\eeq
from which we get the stationary density profile 
\beq 
n_{eq}({\bf r}) = \mu^{-1}\left( \bar{\mu} - U({\bf r}) \right) \; , 
\eeq
where $\mu^{-1}(y)$ is the inverse function of $\mu(n)$. 
This is exactly the equation one finds in the stationary case 
from Eq. (\ref{hy-2}). Obviously this stationary 
density profile strongly depends on the shape of $\mu(n)$, 
i.e. it depends on statistics and interaction strength. 

We can also study small deviations from the equilibrum 
configuration $\Psi_{eq}({\bf r})$ by setting 
\beq 
\Psi({\bf r},t) = \left( \Psi_{eq}({\bf r}) + \phi({\bf r},t) \right) 
\ e^{-i\zeta \bar{\mu} t/\hbar} \; .  
\eeq 
It is strightforward to show that, after 
introducing the sound velocity of the bulk system 
\beq 
c(n) = \sqrt{{n\over m}{\partial \mu(n)\over \partial n}} \; , 
\eeq
from the linearization of Eq. (\ref{super}) 
the perturbation $\phi({\bf r},t)$ satisfies the equation 
\beq 
i \hbar {\partial \over \partial t} \phi({\bf r},t) 
= \left[ -{\hbar^2 \over 2\zeta m} \nabla^2 
- \zeta \bar{\mu} + \zeta U({\bf r}) 
+ \zeta \mu(n_{eq}({\bf r})) + 
\zeta c(n_{eq}({\bf r}))^2 \right] \phi({\bf r},t) 
+ \zeta  c(n_{eq}({\bf r}))^2 \ \phi^*({\bf r},t) \; .   
\label{linear} 
\eeq 
In the uniform case, where $U({\bf r})=0$ and so 
$\bar{\mu}=\mu(n_{eq})$, we set 
\beq 
\phi({\bf r},t) = A e^{i ({\bf k}\cdot {\bf r}-\omega t)} + 
B e^{-i({\bf k}\cdot {\bf r}-\omega t)} \; , 
\eeq
and from Eq. (\ref{linear}) we find the Bogoliubov dispersion relation 
\beq 
\hbar \omega = \sqrt{{\hbar^2k^2\over 2\zeta m}
\left( {\hbar^2k^2\over 2\zeta m} + 2\zeta m c(n_{eq})^2 \right)} \; . 
\label{dispersion}
\eeq
In the long wavelength limit we find the phonon-like spectrum 
\beq 
\omega = c(n_{eq}) \ k \;  
\label{phonon}
\eeq 
of sound waves, while in the short wavelength limit 
we obtain the particle-like spectrum 
\beq 
\hbar \omega = {\hbar^2k^2\over 2\zeta m} \; ,  
\label{single-particle} 
\eeq
that is the kinetic energy of a boson if $\zeta=1$ or 
that of a Cooper-pair if $\zeta=2$. 
Clearly Eq. (\ref{single-particle}) is not reliable 
because the superfluid equations 
are valid only in the low-energy and long wavelength regime. 
Notice that by considering the next-to-leading term 
in the small-momentum expansion of Eq. (\ref{dispersion}) one gets  
\beq 
\omega = c(n_{eq}) \ k + {\hbar^2\over 8 \zeta^2 m^2 \ c(n_{eq})} k^3 \; ,  
\eeq
and this phonon dispersion relation distinguishes between 
superfluid bosons ($\zeta=1$) and superfluid fermions ($\zeta=2$) 
also if the sound velocity $c(n_{eq})$ would be the same. 

In conclusion we stress that the superfluid nonlinear 
Schr\"odinger (SNLS) equation we have introduced can be used 
for both superfluid bosons and fermions in all 
the situations where the characteristic wavelengths of 
the phenomenon under investigation is larger than 
the healing length given by Eq. (\ref{healing}). 
The SNLS equation can be applied to study not only 
collective modes \cite{stringa1,stringa2,kim,manini05} 
and free expansion \cite{stringa1,stringa2,sala-expansion}, but also 
quantized vortices where the characteristic 
length is the vortex-core size \cite{stringa1,sala-ggpe}, 
and tunneling phenomena where the characteristic length 
is the tunneling penetration depth \cite{sala-gle,smerzi}. 
In addition, with the SNLS equation one can 
investigate interesting nonlinear effects, 
like solitons \cite{adhikari,sala-solitons}, 
shock waves \cite{sala-shock}, and also chaos \cite{sala-chaos}. 
Finally, we observe that the SNLS equation satisfies the requirements 
suggested by Greiter, Wilczek and Witten \cite{witten} 
to have a well-founded theory of neutral superconductors: 
it is Galilei invariant, it mantains the current-momentum 
algebric identity, and at low-energy it shows a Nambu-Goldstone 
boson field with linear dispersion relation, 
i.e. the phonon spectrum of Eq. (\ref{phonon}). 

The author has been partially supported by Fondazione CARIPARO 
and GNFM-INdAM. He thanks Flavio Toigo, Nicola Manini, Alberto 
Parola, Boris Malomed and Sadhan Adhikari for many enlightening 
discussions.


\begin{thebibliography}{99}
\bibitem{nota} Note that the nonlinear Schr\"odinger equation recently used 
by Kim and Zubarev \cite{kim} and others \cite{manini05,adhikari} 
to describe a Fermi superfluid has the coefficient 
$\zeta=1$ instead of $\zeta=2$. This choice implies an incorrect 
relationship between phase and superfluid velocity.
\end{thebibliography}
\end{document}